\documentclass[aps,twocolumn,a4paper,10pt]{revtex4-1}
\usepackage{mathrsfs}
\usepackage{graphicx}
\usepackage{latexsym}
\usepackage{amsmath}
\usepackage{amssymb}
\usepackage{textcomp}
\usepackage{amsbsy}
\usepackage[usenames]{color}
\usepackage{epstopdf}

\begin{document}
 \title{\large \bf Bouncing cosmology from warped extra dimensional scenario}
\author{Ashmita Das}
\email{ashmita@iitg.ernet.in}
\author{Debaprasad Maity}
\email{debu@iitg.ernet.in }
\affiliation{Department of Physics, Indian Institute of Technology, North Guwahati, Guwahati, Assam 781039, India}
\author{Tanmoy Paul}
\email{pul.tnmy9@gmail.com}
\author{Soumitra SenGupta}
\email{tpssg@iacs.res.in}
\affiliation{Department of Theoretical Physics,\\
Indian Association for the Cultivation of Science,\\
2A $\&$ 2B Raja S.C. Mullick Road,\\
Kolkata - 700 032, India.\\}

\begin{abstract}
From the perspective of four dimensional effective theory on a two brane warped geometry model,  
we examine the possibility of ``bouncing phenomena''on our visible brane. 
Our results reveal that the presence of warped extra dimension lead to a non-singular bounce 
on the brane scale factor and hence can remove the ``big-bang singularity''. 
We also examine the possible parametric regions for which this bouncing is possible.
\end{abstract}
\maketitle

\section{Introduction}
Over the last two decades models with extra spatial dimensions \cite{arkani,horava,RS,kaloper,cohen,burgess,chodos}  has been increasingly playing 
a central role in physics beyond standard model of particle \cite{rattazzi} and cosmological \cite{marteens} Physics. 
Apart from phenomenological approach, higher dimensional scenarios occur naturally in string theory. Depending on
different possible compactification schemes for the extra dimensions, a large number of models have been constructed, and
their predictions are yet to be observed in the current experiments. In all these models, our visible universe is identified
as one of the 3-branes embedded within a higher dimensional spacetime. The low energy effective description \cite{kanno,shiromizu,sumanta} 
of the dynamical 3-brane turned out to be a very powerful tool in studying dynamics ranging from particle to cosmology. In our
current work we will take this root to understand cosmological bouncing phenomena in the early universe cosmology considering Randall-Sundrum
two brane model.

Among various extra dimensional models proposed over last several years, Randall-Sundrum (RS) warped extra dimensional model \cite{RS} earned a 
special attention since it can resolve the gauge hierarchy problem without introducing any 
intermediate scale (between Planck and TeV) in the theory. RS model is a five dimensional AdS space  
with $S^1/Z_2$ orbifolding along the extra dimension while two 3-branes are placed at the orbifold fixed points. 
The bulk negative cosmological constant along with appropriate boundary conditions 
generate exponentially warped geometry along the extra dimension. Due to this exponential warping, 
the Planck scale on one brane gets suppressed along the extra dimension and emerges as TeV scale \cite{RS} on the visible
brane. In RS model
the interbrane separation (known as modulus or radion) is $\sim$ Planck length and  
generates the required hierarchy between the branes. Subsequently, Goldberger and Wise (GW) proposed a modulus stabilzation mechanism \cite{GW} 
by introducing a massive scalar field in the bulk with appropriate boundary conditions. Different variants 
of RS model and its modulus stabilization are extensively studied in \cite{GW_radion,csaki,julien,ssg1,ssg2,tp1,tp2}.
In this paper we will consider a specific variant of RS scenario and study the cosmological dynamics
from the perspective of low energy effective field theory induced on the visible brane. 

It is well known that standard Big Bang scenario is quite successful in explaining 
many aspects of cosmological evolution of our universe. However, 
the big bang model is plagued with a singularity (known as ``cosmological singularity'') in the
finite past. Resolving this time like cosmological singularity is an important 
issue which is a subject of great research in theoretical cosmology for the last several decades. 
It is widely believed that quantum theory of gravity, if any, should play very important role in resolving this singularity.
One of the important aspects of all the known non-singular cosmological models is the existence of pre Big-Bang universe \cite{gasperini}.
In terms of effective theory, different models of non-singular cosmologies such as 
Ekpyrotic universe \cite{steinhardt}, Loop quantum cosmology \cite{ashtekar}, Galileon genesis \cite{nicolis}, or classical bouncing model, 
can be described by gravity coupled to a scalar field which generically violates null energy condition at the background level. 
Therefore, the scale factor of the universe undergoes a non-singular bounce from a pre-existing universe to the present universe. 
This fact resulted into a reasonable amount of work on classical bouncing cosmology \cite{bc1,bc2,bc3,bc4,bc5,barrow1}, with/without the presence of matter 
components (see also \cite{bc6,bc7,bc8,bc9}).
    
 In the present work, we will study the dynamics of the induced low energy theory 
 which contains modes originating from bulk Physics.
 The dynamics of such mode in the context of usual cosmology \cite{cos1,cos2,cos3,cos4,cos5,cos6} 
has been studied extensively. However, here we ask the following question:

\begin{itemize}
 \item Can the effect of extra dimension trigger a non-singular bounce on the brane scale factor 
 and allow to remove the ``big bang singularity'' ?
\end{itemize}

In the context of two brane scenario, Ekpyrotic model \cite{steinhardt} and its various other variants 
are known to have cosmological bouncing solutions. Important point to emphasize that, 
in the those scenarios the bounce occurs at the time when
two branes collapse. However, in this paper we will be studying the possibility of bouncing phenomena strictly in the
Randall-Sundrum framework, where, gauge hierarchy will impose further restriction on the moduli (radion) dynamics. 
In this regards, we have employed radion stabilization mechanism in the time dependent RS background such that
it does not spoil the bouncing phenomena. Our classical effective field theory computation shows that the 
required gauge hierarchy can be obtained in the asymptotic limit after the bounce.

The aim of this paper is to address aforementioned question in the backdrop of a generalized scenario 
of RS model proposed in \cite{kanno}. The effective on-brane action, we used in this paper, is formulated by Kanno and Soda in \cite{kanno} 
by the method of ``low energy expansion scheme''.\\
Our paper is organized as per the following sequence: in section [II], we briefly describe the 
generalized RS model and its effective action on the visible brane. In section [III], we present the 
cosmological solutions of effective Friedmann equations. The stabilization mechanism of radion field 
is discussed in section [IV] and finally we end the paper with some conclusive remarks.

\section{Low energy effective action on the visible brane}
In RS model, the Einstein equations are derived for a fixed inter-brane separation as 
well as for flat 3-branes. However, the scenario changes if the distance between the branes 
becomes a function of spacetime coordinates and the brane geometry is curved. These 
generalizations are incorporated while deriving the effective action on the TeV brane via 
``low energy expansion scheme'' proposed in \cite{kanno}.

The model we considered in the present paper is described by a five dimensional 
anti-de Sitter (AdS) spacetime with two 3-branes embedded within the spacetime. 
The spacetime geometry has $S^1/Z_2$ orbifolding along the extra dimension. 
Taking $\varphi$ as the extra dimensional angular coordinate, the branes are situated at 
orbifolded fixed points i.e. at $\varphi=0$ (Planck brane) and $\varphi=\pi$ (TeV brane) 
respectively while our visible universe is identified with the TeV scale brane. 
The proper distance between the branes is considered as a function of spacetime coordinates. 
The action of this model \cite{kanno} is the following: 
\begin{eqnarray}
 S&=& \frac{1}{2\kappa^2} \int d^4x d\varphi \sqrt{-G} [R^{(5)}+ (12/l^2)]\nonumber\\ 
 &-&\int d^4x [\sqrt{-g_{hid}} V_{hid} + \sqrt{-g_{vis}} V_{vis}]
 \label{five dim action}
\end{eqnarray}
with $x^\mu=(x^0,x^1,x^2,x^3)$ are the brane coordinates. 
$\frac{1}{2\kappa^2} = M^3$, $M$ is the five dimensional Planck mass. $R^{(5)}$ and 
$l$ ($\sim$ Planck length) are the Ricci scalar and curvature radius of the five dimensional spacetime respectively. 
Critical brane tensions of hidden and visible brane are respectively given by, $V_{hid}$ and 
$V_{vis}$.

We use the following metric ansatz \cite{kanno}, 
\begin{equation}
 ds^2 = b^2(x)d\varphi^2 + e^{-2A(\varphi,x)}h_{\mu\nu}(x)dx^\mu dx^\nu ,
 \label{five dim metric}
\end{equation}
where, $A(\varphi,x)$ is space-time dependent warp factor along the extra dimension, and $b(x)$ is  
radius of the compactified extra dimension. For this above metric ansatz, the five 
dimensional Einstein equations are given by:

\begin{eqnarray}
 &\frac{e^{-2\xi}}{b}&(K^\mu_\nu)_{,\varphi} - e^{-2\xi}KK^\mu_\nu + R^{(4)^\mu}_\nu(h) - \nabla^\mu\nabla_\nu\xi 
 - \nabla^\mu\xi\nabla_\nu\xi\nonumber\\
 &=&-\frac{4}{l^2}\delta^\mu_\nu + \kappa^2(\frac{1}{3}\delta^\mu_\nu V_{hid})
 \frac{e^{-\xi}}{b}\delta(\varphi)\nonumber\\
 &+&\kappa^2(\frac{1}{3}\delta^\mu_\nu V_{vis})\frac{e^{-\xi}}{b}\delta(\varphi-\pi)
 \label{equation1}
\end{eqnarray}

\begin{eqnarray}
 &\frac{e^{-2\xi}}{b}&K_{,\varphi} - e^{-2\xi}K^{\mu\nu}K_{\mu\nu} - \nabla^\mu\nabla_\mu\xi - \nabla^\mu\xi\nabla_\mu\xi\nonumber\\
 &=&-\frac{4}{l^2} + \frac{4\kappa^2}{3}V_{hid}\frac{e^{-\xi}}{b}\delta(\varphi)\nonumber\\ 
 &-&\frac{4\kappa^2}{3}V_{vis} \frac{e^{-\xi}}{b}\delta(\varphi-\pi)
 \label{equation2}
\end{eqnarray}

\begin{eqnarray}
 \nabla_\nu(e^{-\xi}K^\nu_\mu) - \nabla_\mu(e^{-\xi}K) = 0
 \label{equation3}
\end{eqnarray}

where $R^{(4)^\mu}_\nu(h)$ is the Ricci curvature, formed by the metric $h_{\mu\nu}$. 
$K_{\mu\nu}=\frac{1}{b(x)}A'(\varphi)e^{-2A}h_{\mu\nu}$ denotes the extrinsic curvature of $\varphi=$ constant 
hypersurface and $\nabla_{\mu}$ is the covariant 
derivative with respect to $h_{\mu\nu}$. Moreover we introduce $e^{\xi}= \frac{b(x)}{l}\pi$.

In order to solve the five dimensional Einstein equations, it is assumed that  
the brane curvature radius $L$ is much 
larger than the bulk curvature $l$ i.e. $\epsilon= (\frac{l}{L})^2 \ll 1$. Then the bulk Einstein equations 
can be solved perturbatively where $\epsilon$ is taken as the perturbation parameter. This method 
is known as ``low energy expansion scheme'' \cite{kanno} in which the metric is expanded with increasing 
power of $\epsilon$. The zeroth order perturbation solution replicates the RS situation where the inter-brane 
separation is constant. The effective on-brane action can be obtained up to 
first order perturbation, incorporates the fluctuation of modulus as well as non-zero value 
of brane matter. Taking these generalizations into account, the $y$ dependence of the warp factor 
can be obtained as follows (due to Kanno and Soda, see \cite{kanno}):
\begin{equation}
 A(\varphi,x) = \frac{b(x)}{l}\varphi
 \label{warp factor}
\end{equation}

Plugging back the solutions into original five dimensional action (in eqn.(\ref{five dim action})) 
and integrating over the extra dimensional coordinate yields the effective four dimensional 
action for visible brane and given by (see \cite{kanno}),
\begin{eqnarray}
 S_{eff}&=&\frac{l}{2\kappa^2} \int d^4x \sqrt{-f} [\Phi(x)R^{(4)}(f)\nonumber\\ 
 &+&\frac{3}{2(1+\Phi)}h^{\mu\nu}\partial_\mu\Phi\partial_\nu\Phi]
 \label{effective action}
\end{eqnarray}

where $\Phi(x) = [\exp{(2\pi\frac{b(x)}{l}})-1]$ and $R^{(4)}(f)$ is the Ricci scalar formed by 
the induced metric of the visible brane i.e. $f_{\mu\nu}$ ($= e^{-A(\pi,x)}h_{\mu\nu}$).  
It may be noticed from eqn.(\ref{effective action}) that upon 
projecting the bulk gravity on the brane, the extra degrees of freedom of $R^{(5)}$ 
(with respect to $R^{(4)}(f)$) appears as scalar field $\Phi(x)$ which directly couples with 
the four dimensional Ricci scalar. Hence the effective on-brane action 
is a Brans-Dicke like theory.

Eqn.(\ref{five dim metric}) leads to the separation between hidden and visible brane along the path 
of constant $x^\mu$ as follows :
\begin{equation}
 d(x) = \int_{0}^{\pi} d\varphi b(x) = \pi b(x)
 \label{brane separation}
\end{equation}
Above expression (eqn.(\ref{brane separation})) clearly indicates that the proper distance 
between the branes depends on the brane coordinates and thats why 
$d(x)$ can be treated as field. From the perspective of four dimensional effective theory, 
this field is termed as 'radion field' (or modulus field) which is symbolized by $\Phi(x)$ in eqn.(\ref{effective action}).

\section{Cosmological solution for effective on-brane theory}
Considering the effective four dimensional action presented in eqn.(\ref{effective action}), one obtains 
the equations of motion for gravitational and scalar field as follows:
\begin{eqnarray}
 &\Phi&E_{\mu\nu} + f_{\mu\nu}[\Box\Phi + \frac{3}{4(1+\Phi)}\nabla_{\alpha}\Phi\nabla^{\alpha}\Phi]\nonumber\\
 &-&\nabla_{\mu}\nabla_{\nu}\Phi - \frac{3}{2(1+\Phi)}\nabla_{\mu}\Phi\nabla_{\nu}\Phi = 0
 \label{eqn1}
\end{eqnarray}
and
\begin{eqnarray}
 \frac{3}{(1+\Phi)}\Box\Phi - \frac{3}{2(1+\Phi)^2}\nabla_{\mu}\Phi\nabla^{\mu}\Phi = 0
 \label{eqn2}
\end{eqnarray}

where $E_{\mu\nu}$ is the Einstein tensor and the covariant derivatives are formed by the visible brane 
metric $f_{\mu\nu}$. Consider the on-brane metric ansatz as FRW metric with negative curvature parameter,
\begin{eqnarray}
 ds_{(4)}^2&=&f_{\mu\nu}dx^{\mu}dx^{\nu}\nonumber\\
 &=&-dt^2 + a^2(t)[\frac{dr^2}{(1+r^2)} + r^2d\Omega^2]
 \label{metric ansatz}
\end{eqnarray}
where $a(t)$ is the scale factor and $x^{\mu}=(t,r,\Omega)$ are the 
spherical polar coordinates. Using this metric 
ansatz, the field equations (eqn.(\ref{eqn1}) and eqn.(\ref{eqn2})) take the following form:
\begin{equation}
 H^2 = \frac{1}{a^2} - H\frac{\dot{\Phi}}{\Phi} + \frac{(\dot{\Phi})^2}{4\Phi(1+\Phi)}
 \label{eqn3}
\end{equation}
and
\begin{equation}
 \ddot{\Phi} = -3H\dot{\Phi} + \frac{(\dot{\Phi})^2}{2\Phi(1+\Phi)}
 \label{eqn4}
\end{equation}
An overdot denotes $\frac{d}{dt}$, $H=\dot{a}/a$ is known as Hubble parameter and we assume that the radion field 
$\Phi$ is homogeneous in space.\\
In order to solve the above coupled equations (eqn.(\ref{eqn3}) and eqn.(\ref{eqn4})), we 
adopt the procedure formulated in {\cite{barrow2}. Introducing conformal time through 
\begin{equation}
 a d\eta = dt
 \label{conf time}
\end{equation}
and denoting $\frac{d}{d\eta}$ by a prime, eqn.(\ref{eqn4}) becomes - 
\begin{equation}
 \Phi'' + 2\frac{a'}{a} = \frac{(\Phi')^2}{2\Phi(1+\Phi)}
 \label{eqn5}
\end{equation}
Integrating eqn.(\ref{eqn5}), we have the following solution:
\begin{equation}
 \Phi' a^2 = B\sqrt{1+\Phi}
 \label{eqn6}
\end{equation}
where $B$ is a constant. Defining a new variable,
\begin{equation}
 y = \Phi a^2
 \label{petzold variable}
\end{equation}
eqn.(\ref{eqn3}) becomes,
\begin{equation}
 (y')^2 = 4y^2 + \frac{(\Phi')^2 a^4}{1+\Phi}
 \nonumber
\end{equation}
which, along with eqn.(\ref{eqn6}) yields
\begin{equation}
 (y')^2 = 4y^2 + B^2
 \label{eqn7}
\end{equation}

Once the Friedmann equations are expressed in terms of the variable $y$, solutions 
of scale factor ($a(t)$) and radion field ($\Phi(t)$) can be obtained by 
performing the following steps:

\subsection{Step 1: Solution for $y=y(\eta)$}
Integrating eqn.(\ref{eqn7}), we obtain the solution for $y=y(\eta)$ as:
\begin{equation}
 y(\eta) = \frac{1}{2}B \sinh[2(\eta+\eta_0)]
 \label{sol1}
\end{equation}
where $\eta_0$ is an integration constant.

\subsection{Step 2: Solution for $\Phi=\Phi(\eta)$}
Dividing both sides of eqn.(\ref{eqn6}) by $y$, we get the integral of $\Phi(\eta)$ as:
\begin{equation}
 \int \frac{d\Phi}{\Phi\sqrt{1+\Phi}} = B \int \frac{d\eta}{y(\eta)}
 \nonumber
\end{equation}
where we use the definition $y=\Phi a^2$. By putting the solution  of $y(\eta)$ into the 
above equation, one lands up with the following form of $\Phi=\Phi(\eta)$
\begin{equation}
 \Phi(\eta) = \frac{4D \tanh(\eta+\eta_0)}{[1-D \tanh(\eta+\eta_0)]^2}
 \label{sol2}
\end{equation}
with $D$, an integration constant.

\subsection{Step 3: Solution for $a=a(\eta)$}
Plugging the solutions of $y(\eta)$ and $\Phi(\eta)$ into the expression 
$y=\Phi a^2$, the solution  of scale factor with respect to the 
conformal time is found to be:
\begin{equation}
 a^2(\eta) = \frac{B}{4D} [\cosh(\eta+\eta_0) - D \sinh(\eta+\eta_0)]^2
 \label{sol3}
\end{equation}

\subsection{Step 4: Solution for $a=a(t)$ and $\Phi=\Phi(t)$}
From the above solutions of $a(\eta)$, $\Phi(\eta)$ and using eqn.(\ref{conf time}), we 
obtain the scale factor and radion field with respect to cosmic time ($t$) as:
\begin{eqnarray}
 &a(t)&= [t^2 + \frac{B}{4D}(1-D^2)]^{1/2}
 \label{sol4a}\\
 &\Phi(t)&= \frac{D}{(1-D^2)^2}*\nonumber\\
 &[&\frac{32\frac{D^2}{B}t^2 - 8\sqrt{\frac{D}{B}}(1+D^2)t\sqrt{4t^2\frac{D}{B}+(1-D^2)}}{4t^2\frac{D}{B}+(1-D^2)}]
 \label{sol4b}
\end{eqnarray}

It is evident from eqn.(\ref{sol4a}), that $a(t)$ has a non-zero minimum at $t=0$ for $D<1$, where the minimum 
value is given by,
\begin{equation}
 a(0) = [\frac{B}{4D}(1-D^2)]^{1/2}
 \nonumber
\end{equation}
Thus the presence of warped extra dimension allows a non-singular bounce of the scale 
factor (at $t=0$) in our four dimensional universe, as long as the parameter $D$ is constrained to be less than unity.\\
However, it can be checked from eqn.(\ref{sol4b}), that $\Phi(t)$ has a positive asymptotic value as $t\rightarrow -\infty$ 
and goes to zero at $t=\frac{\sqrt{BD}}{2}$. Using eqn.(\ref{sol4b})) 
and the relation $\Phi(t)=[\exp{\big(2\pi\frac{b(t)}{l}\big)} -1]$, we 
obtain figure(1) demonstrating the variation of interbrane separation ($b(t)$) with time.

\begin{figure}[!h]
\begin{center}
 \centering
 \includegraphics[width=3.0in,height=2.0in]{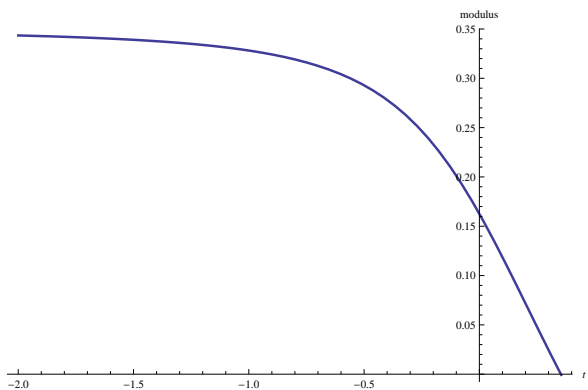}
 \caption{$b(t)$ vs $t$ for $B=1$ and $D=0.5$}
 \label{plot modulus1}
\end{center}
\end{figure}

Figure (\ref{plot modulus1}) clearly 
reveals that the branes collapse into each other within a finite time $t=\frac{\sqrt{BD}}{2}$, 
which indicates the instability 
of the entire set up. Thus we need a suitable mechanism to stablize the modulus. 
Following the procedure 
adopted in \cite{GW,sumanta_radion}, the stabilization method for the present set up is discussed in the next section.


\section{Radion Stabilization}
In order to address the stabilization of time dependent radion field, one needs to consider 
a dynamical stabilization mechanism, which can be achieved by 
time dependent generalization of the Goldberger-Wise (GW) mechanism \cite{GW}. 
Earlier, a similar approach was adopted in \cite{sumanta_radion}. 
Introducing a time dependent scalar field (with quartic brane interactions) in the bulk \cite{GW,sumanta_radion}, 
we address the dynamics of modulus stabilization 
without sacrificing the conditions necessary to resolve the gauge hierarchy problem.\\
The action for the time dependent bulk scalar field is given by,
\begin{equation}
S_5=\frac{1}{2}\int d^4xd\varphi \sqrt{-G}\bigg[G^{MN}\partial_M\Psi\partial_N\Psi + m^2\Psi^2\bigg]\label{actionstable1}
\end{equation}
Where, $M$, $N$ symbolizes $(\mu,\varphi)$. The hidden and visible brane interaction terms with
the bulk stabilizing scalar field can be written as,
\begin{equation}
S_4=\int d^4x \int_{-\pi}^{+\pi}d\varphi \sqrt{-g_h}\lambda_v\bigg[\Psi^2-\tilde{v}_{h}^{2}\bigg]^2 \delta(\varphi-0)\label{actionstable2}
\end{equation}
and
\begin{equation}
S_4=\int d^4x \int_{-\pi}^{+\pi}d\varphi \sqrt{-g_v}\lambda_h\bigg[\Psi^{2} -\tilde{v}_{v}^{2}\bigg]^2 \delta(\varphi-\pi)\label{actionstable3}
\end{equation}
Where, $g_h, g_v$ are the determinant of the metric induced on the hidden and visible brane respectively.
The scalar field action (in eqn. (\ref{actionstable1})) leads to  
the field equation for $\Psi=\Psi(\varphi,t)$ as follows:
\begin{eqnarray}
&\frac{\partial}{\partial t}\bigg[e^{-2A}a^3(t)b(t)\frac{\partial \Psi}{\partial t}\bigg]
-\frac{\partial}{\partial \varphi}\bigg[e^{-4A}\frac{a^3(t)}{b(t)}
\frac{\partial \Psi}{\partial \varphi}\bigg]\nonumber\\
&+m^2 e^{-4A}a^3(t)b(t)\Psi + 4e^{-4A}a^3(t)\lambda_h \psi\bigg(\Psi^2- \tilde{v}_{h}^{2}\bigg)\delta(\varphi)\nonumber\\
&+4e^{-4A}a^3(t)\lambda_v \psi\bigg(\Psi^2- \tilde{v}_{v}^{2}\bigg)\delta(\varphi-\pi)=0\label{scalar1}
\end{eqnarray}
In the limit of large $\lambda_h$ and $\lambda_v$, the boundary conditions for $\Psi(\varphi,t)$ turns out to be,
\begin{equation}
\Psi(0,t)=\tilde{v}_h(t)=F(t)v_h~~~~~~~~~\Psi(\pi,t)=\tilde{v}_v(t)=F(t)v_v\label{bdc1}
\end{equation}
where $F(t)$ carries the time dependence of $\tilde{v}_{h}(t)$ and $\tilde{v}_v(t)$. We choose 
a generalized solution for the stabilizing scalar field as,
\begin{equation}
\Psi(\varphi,t)=F(t)\bigg[P(t)e^{(2+\nu)A} + Q(t)e^{(2-\nu)A}\bigg]\label{scalarsol1}
\end{equation}
where $\nu = \sqrt{4+\frac{m^2}{k^2}}$, $k=\frac{1}{l}$ and recall that $A(\varphi,t) = kb(t)\varphi$. 
Using the boundary conditions we obtain,
\begin{equation}
P(t)=v_ve^{-(2+\nu)k\pi b(t)} - v_h e^{-2\nu k\pi b(t)}\label{coeffecient1}
\end{equation}
and
\begin{equation}
Q(t)=v_h(1+e^{-2\nu k\pi b(t)}) - v_ve^{-(2+\nu)k \pi b(t)}\label{coeffecient2}
\end{equation}
Moreover, using the scalar field solution presented in eqn.(\ref{scalarsol1}), the 
time dependent part of the differential equation (\ref{scalar1}) takes the following form,
\begin{eqnarray}
&a^3(t)b(t)\bigg[e^{\nu A}\bigg\{F(t)P_t + P(t)\bigg(F_t+(\nu+2)F(t)A_t\bigg)\bigg\}\nonumber\\
&+e^{-\nu A}\bigg\{F(t)Q_t+Q(t)\bigg(F_t + (2-\nu)F(t)A_t\bigg)\bigg\}\bigg]\nonumber\\
&=C(\varphi)\label{scalarsol2}
\end{eqnarray}
Where, $F_t, P_t, Q_t, A_t$ are the derivatives of $F, P,Q,A$ with respect to $t$ and 
$C(\varphi)$ is a $\varphi$ dependent integration constant.\\
Plugging back the solutions of $P(t)$ and $Q(t)$ (obtained in eqn.(\ref{coeffecient1}) and eqn. (\ref{coeffecient2})) 
into eqn. (\ref{scalarsol2}), one obtains a differential equation for $F(t)$ as follows:
\begin{equation}
\frac{\partial F}{\partial t}\propto k\frac{e^{2k \pi b(t)}}{a^3(t)}\label{stabilisation2}
\end{equation}
where we consider that the scalar field mass ($m$) is less than the bulk curvature ($k$). Finally,
\begin{equation}
\frac{\partial F}{\partial t}= f_0 k\frac{e^{2k \pi b(t)}}{a^3(t)}\label{stabilisation3}
\end{equation}
with $f_0$, a dimensionless constant. Using the solutions of $a(t)$ and $b(t)$ obtained in eqn.(\ref{sol4a}) and eqn.(\ref{sol4b}), 
the function $F(t)$ can be determined as follows:
\begin{eqnarray}
 &F&(t) =\bigg[f_0k \bigg(\frac{8D\sqrt{\frac{D}{B}}\sqrt{\frac{D}{B}-DB+4t^2}}
 {3(1-D^{2})^3[4D t^2+B(1-D^{2})]^2}\bigg)*\nonumber\\
 &[&4D(3+10D^{2}+3D^{4})+\sqrt{\frac{D}{B}}t^3-3(D^{4}-1)\nonumber\\
 &B&\big\{(1+D^{2})\sqrt{\frac{D}{B}}t
+D\sqrt{1-D^{2}+4\frac{D}{B}t^2}] + E_0\bigg]\label{function1}
\end{eqnarray}
where $E_0$ is an integration constant. After obtaining the explicit form of $F(t)$, GW stabilization mechanism 
can be implemented in this framework, where $\Psi(\varphi,t)$ acts as stabilizing field. 
Minimizing the radion potential, we obtain the value 
of $b_{min}(t)$ as:
\begin{eqnarray}
k\pi b_{min}(t)=4\frac{k^2}{m^2}{\rm ln}\bigg(\frac{v_h}{v_v}\bigg)*F(t)\label{stabilisedradion2} 
\end{eqnarray}
where $b_{min}(t)$ is the stabilized value of the modulus and $F(t)$ is given in eqn.(\ref{function1}). 
It can be checked that $b_{min}(t)$ is positive for all $t$ and has asymptotic values at $t\rightarrow \pm\infty$. Thus 
the branes are never going to be collapsed in presence of the bulk massive scalar field ($\Psi(\varphi,t)$)
and the stabilized interbrane separation (i.e. $b_{min}(t)$) acquires a saturated value at large time.\\
\subsection*{Determination of the constants: $f_0$ and $E_0$}
The solution of $F(t)$ (in eqn.(\ref{function1})) immediately leads the asymptotic values of $b_{min}(t)$ as follows:
\begin{eqnarray}
 &k&\pi b_{min}(t\rightarrow -\infty) = 4\frac{k^2}{m^2}{\rm ln}\bigg(\frac{v_h}{v_v}\bigg)\nonumber\\
 &\bigg[&-f_0k\bigg(\frac{4D}{3B}\bigg)
\frac{(1+3D^{2})(3+D^{2})}{(1-D^{2})^3} + E_0\bigg]\label{stabilisedradion3}
\end{eqnarray}
and 

\begin{eqnarray}
 &k&\pi b_{min}(t\rightarrow \infty) = 4\frac{k^2}{m^2}{\rm ln}\bigg(\frac{v_h}{v_v}\bigg)\nonumber\\
 &\bigg[&f_0k\bigg(\frac{4D}{3B}\bigg)
\frac{(1+3D^{2})(3+D^{2})}{(1-D^{2})^3} + E_0\bigg]\label{stabilisedradion4}
\end{eqnarray}

Now, the constants ($f_0$ and $E_0$) can be determined by equating these asymptotic values 
(as shown in eqn.(\ref{stabilisedradion3}) and eqn.(\ref{stabilisedradion4})) with that obtained in eqn.(\ref{sol4b}) 
(for $t\rightarrow -\infty$) and with GW result (for $t\rightarrow +\infty$), as follows :
\begin{itemize}
 \item $k\pi b_{min}(t\rightarrow -\infty) = \ln{\bigg(\frac{1+D}{1-D}\bigg)}$, obtained from eqn.(\ref{sol4b}).
 
 \item $k\pi b_{min}(t\rightarrow +\infty) = 4\frac{k^2}{m^2}{\rm ln}\bigg(\frac{v_h}{v_v}\bigg)$ (in 
 consonance with GW result \cite{GW}).
\end{itemize}
With the help of above two conditions, one finds the constants $f_0$ and $E_0$ (in terms of $B$, $D$) as follows,

\begin{equation}
 E_0 = \bigg[\frac{1}{2} + \frac{m^2/8k^2}{\ln(v_h/v_v)}\ln\bigg(\frac{1+D}{1-D}\bigg)\bigg]\label{findingE1}
\end{equation}
and
\begin{eqnarray}
f_0 = &\bigg(\frac{1}{k}\bigg)\frac{(1-D^{2})^3}{(3+D^{2})(1+3D^{2})}\nonumber\\
&\bigg[1-\frac{m^2/8k^2}{\ln(v_h/v_v)}{\rm ln}\bigg[\frac{1+D}{1-D}\bigg]\bigg]\label{findingf0}
\end{eqnarray}

Using the above expressions of $f_0$, $E_0$ and eqn.(\ref{stabilisedradion2}), we obtain 
figure(2) between the stabilized modulus ($\frac{b_{min}(t)}{b_{GW}}$) versus $t$ (where $k\pi b_{GW}=4\frac{k^2}{m^2}\ln(v_h/v_v)$):
\begin{figure}[!h]
\begin{center}
 \centering
 \includegraphics[width=3.5in,height=2.8in]{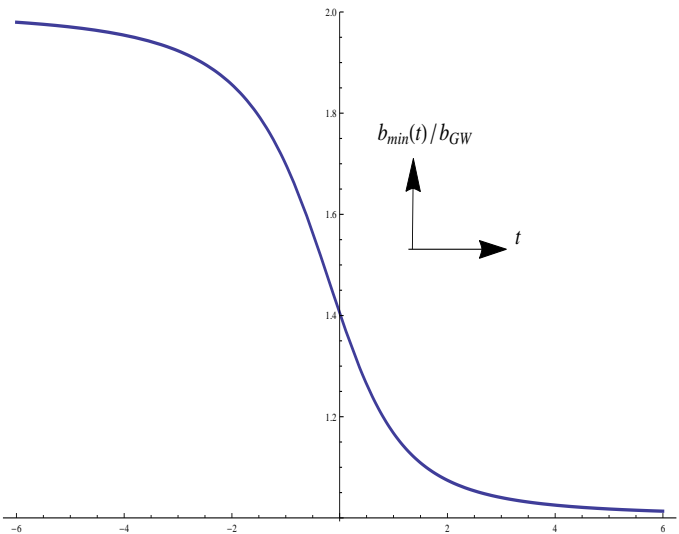}
 \caption{$b_{min}(t)$ vs $t$ for $B=1$ and $D=0.5$}
 \label{plot modulus2}
\end{center}
\end{figure}

Figure(\ref{plot modulus2}) clearly depicts that $b_{min}(t)$ is non-vanishing for the entire range of $t$ ($-\infty <t< \infty$) 
and saturates at 
the Goldberger-Wise value ($b_{GW}$) at large time. Thus the time dependent modulus can be stabilized by 
imposing a time dependent massive scalar field in the bulk. Moreover, we fix the integration 
constants ($f_0$, $E_0$) in such a way that the solution of gauge hierarchy problem is ensured.\\

However the question may arise that whether the introduction of stabilizing scalar field can 
affect the bouncing phenomena or not. To examine this, we substitute the solution of stabilized modulus 
(i.e. $b_{min}(t)$) into the effective Freidmann equation and find,
\begin{eqnarray}
 \big(\frac{\dot{a}}{a}\big)^2&=&\frac{1}{a^2} - 8\frac{k^2}{m^2}{\rm ln}\bigg(\frac{v_h}{v_v}\bigg)\dot{F}(t)
 \frac{\dot{a}}{a}[\frac{\big(\frac{v_h}{v_v}\big)^{8\frac{k^2}{m^2}F(t)}}{\big(\frac{v_h}{v_v}\big)^{8\frac{k^2}{m^2}F(t)}-1}]\nonumber\\
 &+&\bigg(8\frac{k^2}{m^2}{\rm ln}\bigg(\frac{v_h}{v_v}\bigg)\dot{F}(t)\bigg)^2 
 \frac{\big(\frac{v_h}{v_v}\big)^{8\frac{k^2}{m^2}F(t)}}{\big(\frac{v_h}{v_v}\big)^{8\frac{k^2}{m^2}F(t)}-1}
 \label{hubble_stabilization}
\end{eqnarray}
Using the form of $F(t)$ given in eqn.(\ref{function1}), we solve the Hubble parameter ($=\dot{a}/a$) 
numerically and compare this numerical solution with the Hubble parameter obtained earlier (in absence 
of $\Psi(\varphi,t)$, see eqn.(\ref{sol4a})). This comparison is shown in figure(3).

\begin{figure}[!h]
\begin{center}
 \centering
 \includegraphics[width=3.2in,height=2.6in]{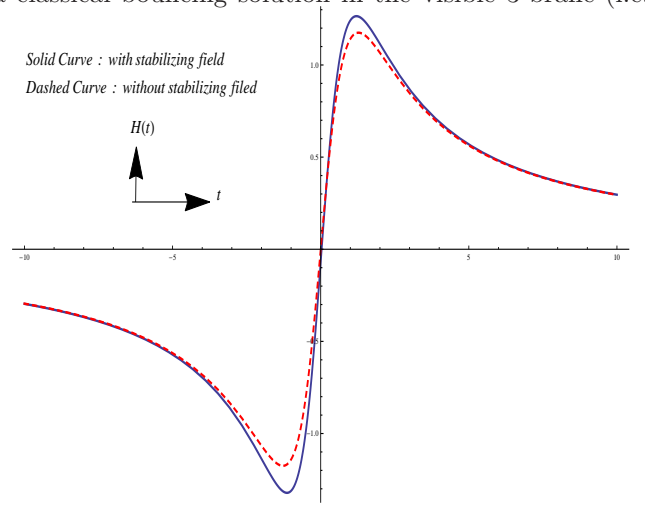}
 \caption{$H(t)$ vs $t$, with/without the stabilizing field for $B=1$, $D=0.5$, $\frac{v_h}{v_v}=1.5$ and $\frac{m}{k}0.2$.}
 \label{plot comparison}
\end{center}
\end{figure}

Figure (\ref{plot comparison}) clearly demonstrates that the feature of the bouncing phenomena 
remains unaffected due to the effect of the stabilizing scalar field.\\

\section{Conclusion}
We consider a five dimensional AdS compactified warped geometric model with 
two 3-branes residing at the orbifold fixed points. Our universe is identified with 
the visible brane. Instead of considering the five dimensional dynamics of brane
under gravity, we studied low energy effective theory induced on our brane following the reference \cite{kanno}. 
In the high bulk curvature limit, the induced four dimensional effective theory appeared to be a  
Brans-Dicke type theory where the scalar field is playing the role of distance modulus 
between the two branes.
In this paper, we investigate the possibility of having a classical bouncing solution in the visible 
3-brane (i.e. our universe). Out of three possible spatial curvature of the
Freedman-Roberston-Walker brane, the bouncing solution exists only for hyperbolic spatial curvature $(\kappa = -1)$. 
Following the procedure as mentioned in section III, it can be shown easily that for $\kappa= 0$ and for $\kappa= +1$, 
one can not have any bouncing solution. 
While finding the solution for $\kappa =-1$, we also introduce the stabilazation mechanism to make sure that the two branes do not collapse, and
maintain the hierarchy of scale in the asymptotic limit.
In addition, we also need to satisfy a specific constraint $D< 1$ to ensure the real valued bouncing solution for the scale factor.\\
As the solution of radion field presented in eqn.(\ref{sol4b}) clearly implies that in an epoch after 
the bouncing, (depicted in figure(\ref{plot modulus1})), the two branes would collapse leading to instability. 
Therefore, in order to stabilize this, a time dependent massive scalar field is introduced in the bulk. 
Thus we have a dynamical stabilization of RS model, where in the asymptotic past the hierarchy of 
scale was larger than that of the present Goldberger-Wise value which is achieved in the asymptotic 
future. This is clearly demonstrated in figure(\ref{plot modulus2}). 
We have determined the stabilization condition in eqn.(\ref{stabilisation2}), and finally
taking this into account, we numerically solve the Hubble parameter as shown in figure (\ref{plot comparison}).
This clearly reveals that the ``bouncing'' phenomena is not affected by the 
stabilizing scalar field.


\begin{thebibliography}{90}
\bibitem{arkani}
N. Arkani-Hamed, S. Dimopoulos, G. Dvali, Phys. Lett.
B 429 263 (1998); N. Arkani-Hamed, S. Dimopoulos, G.
Dvali, Phys. Rev. D 59 086004 (1999); I. Antoniadis, N.
Arkani-Hamed, S. Dimopoulos, G. Dvali, Phys. Lett. B
436 257 (1998)


\bibitem{horava}
P. Horava and E. Witten, Nucl. Phys. B475, 94 (1996); B460, 506 (1996)

 \bibitem{RS}
L. Randall and R. Sundrum, Phys. Rev. Lett. {\bf 83}, 3370 (1999);

\bibitem{kaloper}
 N. Kaloper, Phys. Rev. D60, 123506 1999; T. Nihei,Phys. Lett. B465, 81 (1999); H. B. Kim and H. D. Kim,Phys. Rev. D61, 064003 (2000)
 
 \bibitem{cohen}
  A. G. Cohen and D. B. Kaplan, Phys. Lett. B470, 52(1999); 
  
  \bibitem{burgess}
  C. P. Burgess, L. E. Ibanez, and F. Quevedo,ibid. 447, 257 (1999); 
  
  \bibitem{chodos}
  A. Chodos and E. Poppitz, ibid.471, 119 (1999); T. Gherghetta and M. Shaposhnikov,Phys. Rev. Lett.85, 240 (2000)
  
\bibitem{rattazzi} G. F. Giudice, R. Rattazzi and J. D. Wells, Quantum gravity and extra dimensions at high-energy colliders
Nucl. Phys. B {\bf 544}, 3 (1999). 

\bibitem{marteens} R. Marteens and K. Koyama, Brane-World Gravity, Living Rev. Rel. {\bf 13}, 5 (2010).
  
  \bibitem{kanno}
  S. Kanno and J. Soda, Phys. Rev. D 66, 083506 (2002)
  
  \bibitem{shiromizu}
  T. Shiromizu, K. Maeda, and M. Sasaki, Phys. Rev. D 62, 024012 (2000).
  
  \bibitem{sumanta}
  S. Chakraborty, S. SenGupta,   Eur.Phys.J. C75  11, 538 (2015)
  
  \bibitem{GW} W. D. Goldberger and M. B. Wise, Phys.Rev.Lett.{\bf 83}, 4922 (1999).
  
  \bibitem{sumanta_radion}
  S. Chakraborty, S. SenGupta,  Eur.Phys.J. C74 no.9, 3045 (2014)
  
  \bibitem{GW_radion} W. D. Goldberger and M. B. Wise, Phys.Lett B 475 275-279 (2000)
  
\bibitem{csaki} C. Csaki, M. L. Graesser and Graham D. Kribs,  Phys.
Rev.D.{\bf 63}, 065002.

\bibitem{julien}
J. Lesgourgues, L. Sorbo, Goldberger-Wise variations: Stabilizing brane models with a bulk scalar, Phys. Rev. D69  084010 (2004)

\bibitem{ssg1}
  S. Das, D. Maity, and S. SenGupta, Cosmological constant,
brane tension and large hierarchy in a generalized Randall-
Sundrum braneworld scenario, J. High Energy Phys. {\bf 05}, 042 (2008).

\bibitem{ssg2}
  S. Anand, D. Choudhury, Anjan A. Sen, S. SenGupta, "A Geometric Approach to Modulus Stabilization"\\
   Phys.Rev. D92 (2015) no.2, 026008 (2015); arXiv:1411.5120.
   
   \bibitem{tp1}
A. Das, H. Mukherjee, T. Paul and S. SenGupta, Radion stabilization in higher
curvature warped spacetime,  arXiv:1701.01571 [hep-th].

\bibitem{tp2}
T. Paul,  Brane localized energy density stabilizes the modulus in higher dimensional warped spacetime, 
 arXiv:1702.03722.
 
\bibitem{gasperini} M. Gasperini and G. Veneziano. The Pre - big bang scenario in string cosmology.
Phys.Rept., 373:1–212, (2003).  


\bibitem{steinhardt} J. K. Erickson, {\it et al}, Kasner
and mixmaster behavior in universes with equation of state w ≥ 1. Phys. Rev. D {\bf 69}, 063514 (2004);  
D. Garfinkle, {\it et al}, Evolution to a smooth universe in an ekpyrotic contracting phase with w > 1,
Phys. Rev. D {\bf 78}, 083537 (2008).


\bibitem{ashtekar} A. Ashtekar and P. Singh, Loop Quantum Cosmology: A Status
Report, Class. Quant. Grav. {\bf 28}, 213001 (2011); M. Bojowald, Quantum Cosmology: Effective Theory,
Class. Quant. Grav. {\bf 29}, 213001 (2012). 

 	
\bibitem{nicolis} P. Creminelli, {\it et al},  Galilean Genesis: An Alternative to inflation,
JCAP {\bf 1011}, 021 (2010).    
 
 \bibitem{bc1}
 K. Bamba, A.N. Makarenko, A.N. Myagky, S. Nojiri, S.D. Odintsov,  Bounce cosmology from F(R) gravity and 
 F(R) bigravity, JCAP01 008 (2014)
 
 \bibitem{bc2}
 J. Garriga, A. Vilenkin and J. Zhang, Non-singular bounce transitions in the multiverse,
JCAP 11  055 (2013); [arXiv:1309.2847].

\bibitem{bc3}
B. Gupt and P. Singh, Non-singular AdS-dS transitions in a landscape scenario,
arXiv:1309.2732.

\bibitem{bc4}
Y.-S. Piao, Can the universe experience many cycles with different vacua?, Phys. Rev. D 70
 101302, (2004) [hep-th/0407258].

\bibitem{bc5}
M. Bouhmadi-Lopez, J. Morais and A.B. Henriques, Smoking guns of a bounce in modified
theories of gravity through the spectrum of the gravitational waves, Phys. Rev. D 87 
103528, (2013); [arXiv:1210.1761].

\bibitem{barrow1}
J. D. Barrow; Phys.Rev. D48 3592-3595 (1993)

\bibitem{bc6}
R.H. Brandenberger, The Matter Bounce Alternative to Inflationary Cosmology,
arXiv:1206.4196.

\bibitem{bc7}
M. Novello and S.P. Bergliaffa, Bouncing Cosmologies, Phys. Rept. 463, 127, (2008)
[arXiv:0802.1634].

\bibitem{bc8}
V. Belinsky, I. Khalatnikov and E. Lifshitz, Oscillatory approach to a singular point in the
relativistic cosmology, Adv. Phys. 19,  525, (1970)

\bibitem{bc9}
Y.-F. Cai, D.A. Easson and R. Brandenberger, Towards a Nonsingular Bouncing Cosmology,
JCAP 08, 020, (2012); [arXiv:1206.2382].

\bibitem{cos1}
C. Csaki, M. Graesser, L. Randall, J. Terning,  Phys.Rev. D62, 045015, (2000)

\bibitem{cos2}
P. Binetruy, C. Deffayet, and D. Langlois, Nucl. Phys. B565,
269, (2000͒).

\bibitem{cos3}
C. Csa ́ki, M. Graesser, C. Kolda, and J. Terning, Phys. Lett. B
462, 34 (1999͒).

\bibitem{cos4}
J.M. Cline, C. Grojean, and G. Servant, Phys. Rev. Lett. 83,
4245 (1999͒).

\bibitem{cos5}
P. Kanti, I.I. Kogan, K.A. Olive, and M. Pospelov, Phys. Lett.
B 468, 31 (1999͒).

\bibitem{cos6}
D.J. Chung and K. Freese, Phys. Rev. D 61, 023511 (2000).

\bibitem{barrow2}
J. D. Barrow; Phys.Rev. D47 5329-5335 (1993)
\end{thebibliography}
\end{document}